1    Face processing limitation to own species in primates, a response to social needs?

3    DUFOUR V. [1,2], PASCALIS O. [2] & PETIT O. [1]

4    [1]CEPE, UPR 9010, CNRS, University of Strasbourg, France

5    [2]Department of Psychology, University of Sheffield, UK

7    Face recognition in primates, Dufour et al.

13   Corresponding author: Valérie Dufour, Equipe Ethologie des Primates, 7 rue de l'Universite,

14   67000 Strasbourg, valerie.dufour@neurochem.u-strasbg.fr

15   Pascalis Olivier, Dpt of Psychology, University of Sheffield, S102TP, Sheffield, UK

16   Petit Odile, Equipe Ethologie des Primates, 7 rue de l'Universite, 67000 Strasbourg, France.

18   Text word count: 5821

Most primates live in social groups which survival and stability depends on individuals' abilities to create strong social relationships with other group members. The existence of those groups requires to identify individuals and to assign to each of them a social status. Individual recognition can be achieved through vocalizations but also through faces. In humans, an efficient system for the processing of own species faces exists. This specialization is achieved through experience with faces of conspecifics during development and leads to the loss of ability to process faces from other primate species. We hypothesize that a similar mechanism exists in social primates. We investigated face processing in one Old World species (genus *Macaca*) and in one New World species (genus *Cebus*). Our results show the same advantage for own species face recognition for all tested subjects. This work suggests in all species tested the existence of a common trait inherited from the primate ancestor: an efficient system to identify individual faces of own species only.

There are several lines of evidence that the face processing system presents similarities between different species. First, faces represent a highly attractive stimulus for infant primates, including humans (*Homo sapiens*) (Goren et al. 1975), pigtailed macaques (*Macaca nemestrina*) (Lutz et al. 1998), gibbons (*Hylobates agilis*) (Myowa-Yamakoshi & Tomonaga 2001) and chimpanzees (*Pan troglodytes*) (Kuwahata et al. 2004). Three-week-old lambs (*Ovis ovis*) spend more time sniffing a picture of sheep than a picture displaying other stimuli (Porter & Bouissou 1999). Second, in most primate species, faces are a way to communicate emotions via the production of mimics (van Hoof 1967; Parr 2003; Niedenthal et al. 2000). Third, neurons responding specifically to faces compared to other stimuli have been found in non human primates (Perrett et al. 1992) and in sheeps (Kendrick et al. 2001). In human a face specific electrophysiological response, named the N170, is elicited by faces (Bentin et al. 1996; de Haan et al. 2002). Such similarities may reflect a common origin selected by evolution (Campbell et al. 1997). In the present study we investigate if the advantage of face processing to own species found in humans (Pascalis et al. 2002; Dufour et al. 2004) and in rhesus macaques (*Macaca mulatta*) (Pascalis & Bachevalier 1998) can be found in 2 other non human primate species, Tonkean macaques (*Macaca tonkeana*) and brown capuchin monkeys (*Cebus apella*), which will give convincing evidence to a common origin of the face recognition system.

We shall first review the similarities between the face processing system in human and non human primates. Even if the morphological variability between human faces is limited, human adults are very efficient at recognizing faces; they are able to discriminate hundreds of them (Bahrick et al. 1975). It has been suggested that we differentiated individuals' faces on the basis of relational information, such as the particular distance between the eyes, or between lips and chin (Leder & Bruce 2000). The ability to process these parameters is called *configural* processing and contrasts with feature processing (Diamond & Carey 1986; Maurer et al. 2002). Yin (1969) has shown that faces are recognized more accurately and faster when presented in their canonical orientation than when presented upside-down. A widely accepted claim is that the cause of the inversion effect is a disruption of configural processing leading to the use of a less efficient feature processing (Diamond & Carey 1986; Leder & Bruce 2000; Maurer et al. 2002). Valentine's (1991) model suggests that faces are encoded as individual points within a multidimensional face space defined by a set of dimensions (gender, eye color, etc.). Valentine proposes a norm-based coding model, whereby faces are encoded as vectors according to their deviation from the central tendency, or prototypical average of the face space. It can explain the other-race effect that is that adults find it easier to differentiate faces from their own-ethnic group (Meissner & Brigham 2001). The human face processing system doesn't however extend toward faces of other primate species that share the same face configuration (Dufour et al. 2004; Pascalis et al. 2002).

Conspecifics' face identification is present in non human primates and can be studied using pictures (Pascalis et al. 1999, for a review). For example, Dasser (1987) showed that longtailed macaques could associate the picture of a member of their group to the real individual. Individual's faces discrimination based on pictures has been demonstrated in chimpanzees (Parr et al. 1998), Japanese macaques (*Macaca fuscata*) (Tomonaga 1994) and rhesus macaques (Wright & Roberts 1996; Pascalis & Bachevalier 1998; Parr et al. 1999) among other species. It is important to highlight that those abilities are not restricted to primates but can be observed in other species, such as sheep. Sheeps can categorise their species based on faces (Kendrick et al. 1995). They can also discriminate sheeps of their own breed from sheeps of other breed, and between two individuals from their own breed as measured with a two-way discrimination task (Kendrick et al. 1996).

Similarities in the face processing system include also the inversion effect. Face recognition is sensitive to inversion in sheep (Kendrick et al. 1996) but its presence is still debated for non human primates. In chimpanzees, Tomonaga et al. (1993) did not observe an inversion effect for chimpanzees or humans familiar faces but they tested only one subject with a limited number of stimuli. More recently, Parr et al. (1998) demonstrated an inversion effect in 5 chimpanzees, for human and chimpanzee faces but not for capuchin faces or cars. In longtailed macaques, Bruce (1982) did not observe an effect of inversion using a concurrent discrimination task. Overman and Doty (1982), using a matching-to-sample task, found an inversion effect in pigtailed macaques for both human and macaque faces. An inversion effect was also demonstrated in a squirrel monkey (*Saimiri sciureus*) by Phelps and Roberts (1994) for human faces (but not for monkey faces). The discrepancy observed in the non human primate literature might be due to several factors such as the experimental paradigms that are different. However, since face recognition in sheep is sensitive to inversion, the possibility of the existence of inversion effect in monkeys cannot be rejected.

Nelson (2001) has suggested that the systems underlying human face processing may be sculpted by experience with different kinds of faces present in the visual environment. The face processing system is indeed developing until teenage hood (Carey & Diamond 1994; Campbell et al. 1999). The early system is able to cope with different type of faces, for example, Pascalis et al. (2002) showed that 6-month-old humans discriminated between both human and monkey faces, but that 9-month-olds and adults discriminated only between human faces. Many researchers attribute the other-race and other-species effects to the relatively common experience of having greater exposure to faces of ones own-race compared to other races and greater experience with faces within ones own species compared to other species (Valentine 1991; Nelson 2003). Thus we are best at recognizing faces similar to those we see most often in our environment. However, it is important to differentiate between *other-race faces*, which belong to the same face category as own-race faces (i.e. human faces), and *other-species faces,* which belong to a separate face category (i.e. non human primate). While the face processing system remains flexible for the category of faces to which we are most exposed, humans, this plasticity may not extend for other face categories. Human adults process human faces at an individual level whereas they process other species (monkey, sheep, etc.) at a category level. Experience with other species during infancy seems to be important for the development of the face processing system. In sheep, Kendrick et al. (1998) showed that male sheep reared by goats

(*Capra aegagrus hircus*) and male goats reared by sheep preferred to socialize with females from their maternal species compared with their genetic species. In non human primates, visual preference for maternal species has also been shown for chimpanzees reared by humans (Tanaka 2003) and for Japanese monkeys reared by rhesus macaques (Fujita 1990, albeit not true for rhesus macaques). Pascalis et al. (2005) demonstrate that exposure to Barbary macaques faces (*Macaca Sylvanus*) - between 6 and 9 months of age- facilitates the discrimination of monkey faces, in human infants, an ability that is otherwise lost around 9 months of age. The adults' human limitation to process own-species faces has been observed in non human primates (Pascalis & Bachevalier 1998) and sheep (Kendrick et al. 1996). However, Pascalis and Bachevalier (1998) did not test for the abilities of rhesus macaques to recognize other non human primate species that might have been morphologically closer than human faces. There is indeed evidence that non human primates can discriminate faces of other species. Parr et al. (1998) showed that chimpanzees could learn to discriminate between faces of humans, of their own species, of brown capuchins and between pictures of cars in a sequential matching-to-sample task. Rhesus macaques tested with the same paradigm (Parr et al. 1999) revealed similar ability albeit following a longer training phase. Wright and Roberts (1996) using a two-alternative forced choice task, showed that rhesus macaques could discriminate between human and various species of monkey faces. Other studies confirmed, however, the limitation to process own-species faces. Humphrey (1974) with a habituation dishabituation task showed that rhesus monkeys discriminated between two conspecifics but not between two individuals of another species. Kim et al. (1999), using a visual paired comparison task showed that young pigtailed macaques discriminated faces of their own species but not humans or longtailed macaques faces. From this review, the existence of a limitation to own species in other non human primates remains to be explored. It is indeed not clear whether the face system could extend to the processing of phylogenetically close species faces, process that could be based on morphological proximity of those faces.

Our aim was to study inter-individual discrimination of faces of different species in a wider range of primates with the same technique as the one used by Pascalis and Bachevalier (1998). We investigated face recognition in humans, brown capuchins and Tonkean. Each species tested had none or limited experience with other primate species. In order to test the common origin of the face system, we tested the face processing ability of one macaque species that radiated a few million years before the anthropoid branch, but also of brown capuchin monkeys who diverged earlier than macaques from the ancestral lineage (Fig. 1). Moreover, we explored if the recognition ability extended to phylogenetically close species *i.e.* same-genus species, or was restricted to the recognition of their own species. Therefore, the subjects were tested with faces of more or less close species. We used the visual paired comparison (VPC) task which indexes the relative interest in a pair of visual stimuli consisting of one novel item and one familiar item viewed during a prior familiarization period (Pascalis & de Haan 2003). Recognition is inferred from the participant's tendency to fixate more towards one stimulus than the other; usually the novel stimulus. This task can be used with both human and non human primates.

The VPC task is rather an unusual task to study face recognition in non human primates. It is mainly used to study cognition in both human and non human primate infants but has been recently used successfully to study the neural substrate of recognition memory in monkeys. The VPC has been found to be more sensitive to detect memory deficits after cortical lesions than classic matching tasks (Pascalis & Bachevalier, 1999; Zola et al., 2000; Nemanic et al., 2004).

MATERIAL AND METHODS

Subjects

Five men and four women aged from 23 to 55 years (mean 32 years) were tested at the University of Strasbourg. None of the participants had previous experience with monkeys. Non human subjects were five Tonkean macaques aged from 3 to 11 years old (three females and two males) and five brown capuchin monkeys aged from 3 to 29 years (three females and two males). Subject were all mother reared in social groups. One Tonkean macaque (male) had been isolated for social reasons in an individual cage 1 year before the beginning of the experiment. The other Tonkean macaques were socially housed in an indoor cage of 20-meter square composed of two compartments, connected to a tunnel leading to the testing room. Capuchin monkeys belonged to a social group of 18-22 individuals leaving in an indoor-outdoor enclosure composed of 78 meter square in total (33 meter square inside, 45 meter square outside). Cages and enclosure of all subjects were furnished with wooden perches and plastic elements.

All subjects had an ad libitum access to commercial monkey diet and water. In addition, they received fresh fruit supply once a week. They were never deprived in food or water.

Stimuli

Humans performed 2 trials per category of stimuli. Each trial consisted in a familiarization period with a face followed by a recognition test during which two faces, one new and one being the familiar seen previously were presented. Participants performed 2 trials with human faces, 2 trials with rhesus macaques faces and 2 trials with Tonkean macaques faces Human were tested in order to replicate previous results that showed species specificity in humans (Pascalis & Bachevalier, 1998; Pascalis et al., 2002) with a different set of stimuli.

Non-human primates were tested with four categories of faces, with 10 trials per category. Tonkean macaques were tested with faces of their own species, of stumptailed macaques (*Macaca arctoides*) of rhesus macaques and of humans. Brown capuchin monkeys were tested with faces of their own species, of white-faced capuchins (*Cebus capucinus*) that belong to the same genus, of longtailed macaques and of humans. Stimuli were black and white pictures of unfamiliar individuals. For human as well as for non human primates, contrast and brightness were visually adjusted so as to make pictures appear as similar as possible. It was achieved by controlling the level of contrast and brightness of each paired stimuli using the software Adobe*Photoshop*. Macaques' faces were extracted from Dr B. Thierry private's collection. Capuchin pictures were taken in various French zoological parks, with semi-free individuals that could be seen from a close distance and without mesh. Faces were selected for their neutral expression. They were paired with individual of the same sex. Only frontal view or faces deviating slightly from frontal view were selected. Faces were consequently paired so as to match in their orientation. Figure 2 shows an illustration of the stimuli used.

Procedure

*Testing*

The procedure used for humans is comparable as the one described by Pascalis et al. (2002). Participants were asked to relax and passively view the images that would appear on the screen approximately 60 cm away in front of them. A black and white CCD camera filmed the participant's eye movements for further analysis. Time was recorded and displayed on the control monitor using an Horita (TG-50) at 25 frames per second. The experimenters predetermined a series of images for participant. Each trial consisted of a familiarization period during which a single image was displayed for 3 s from the first time of looking, followed by a blank interval of 3 s, followed by the recognition test that consisted in the presentation of two images, one novel and one familiar, for 3 s. The left-right position of the novel and familiar stimuli was counterbalanced across the testing. The testing of human participants was done over two days. Non-human primates were first isolated from the group and trained to look at the projection screen distant to the cage from one meter. Monkeys were given orange juice each time they looked toward the screen to maintain their attention. Sessions ended as soon as the level of interest of the subject for the experiment declined. One session lasted generally for 4 to 12 trials. Subjects were tested daily before returning into their social group. Training and testing required about 3 months per individual. The Tonkean macaques came from different groups and their testing was done between June 1999 and January 2001. Capuchin monkeys testing began in June 2002 and lasted until March 2003. For non human primates, the familiarization period ended when each stimulus was presented for a total cumulated time of 3 s. The interval lasted 3 s. The recognition test was 3 s of cumulated time. It insured that each participant (whatever the species) watched the stimuli for a similar amount of time. The experimenter estimated the total cumulated time by observing the subjects' behaviour through the monitor. The familiarization time was later controlled picture by picture using the software *the observer*. Using the same software, participant's behaviour during the preference test was then analyzed by measuring time spent looking on the left or on the right stimulus. This was assessed by tracking on videos the corneal reflection of the slide (Fig. 3). Trials were scored by an experimenter, who was not aware of the position of the familiar and novel stimuli.

Data analysis

Statistical analyses were realized on real looking time in humans. In non human primates, the amount of time spent exploring the pictures varied widely between the species, so the data were normalized. For the three species tested, analyses of variance with repeated measures were conducted on the amount of time spent looking at the stimuli considering the following factors: subject, direction of gaze (towards the new stimulus and the familiar) and category of stimuli. Data were further analyzed per category of stimuli, with one-tailed paired t-tests.

RESULTS

For humans, an analysis of variance of the time spent looking at each face (the new and the familiar) reveals an effect of the direction of the gaze ($F_{1, 8} = 6.03$; $P = 0.04$), participants spent more time looking at the novel pictures. Detailed analysis for each category of face shows that this novelty preference is statistically significant only for human faces as revealed by one tailed paired t-test ($T = 2.34$, $df = 8$, $P = 0.03$). There is no evidence of preference for novel or familiar stimuli for faces of another primate species (one tailed paired t-tests for rhesus macaque, $T = 1.31$, $df = 8$, $P = 0.11$; for Tonkean macaques, one tailed paired $T = 0.64$, $df = 8$, $P = 0.27$) (Fig. 4). This result replicates previous results showing that recognition of faces, as measured with a VPC task, is limited to own species in human adults (Pascalis & Bachevalier 1998; Pascalis et al. 2002).

Tonkean macaques were shown paired faces of their own species, of rhesus macaques, of stumptailed macaques and of humans. Results of the analysis of variance reveal an effect of the species ($F_{3, 12} = 5.88$; $P = 0.0008$). Subjects looked longer at Tonkean macaques' faces compared to faces of another species (contrast post-hoc test: Tonkean macaques versus others: $P = 0.0183$). A one tailed paired t-test shows that subjects looked longer at the new stimulus when tested with faces of their own species ($T = 1.73$, $df = 49$, $P = 0.04$). This novelty preference is not found for faces of stumptailed macaques (one tailed paired $T = 1.21$, $df = 49$, $P = 0.12$), rhesus macaques (one tailed paired $T = 0.65$, $df = 49$, $P = 0.26$), or humans (one tailed paired $T = 0.99$, $df = 49$, $P = 0.16$) (Fig. 4). As humans, Tonkean macaques present visual recognition only for their own species. It extends the results obtained with rhesus macaques by Pascalis and Bachevalier (1998).

Concerning brown capuchin monkeys, the analysis of variance reveals that subjects looked longer when faces of their own species were shown, compared to other species (main effect: species, $F_{3,12} = 5.21$, $P = 0.002$; contrast post hoc test, brown capuchin monkeys versus the others: $P = 0.0196$). Novelty preference was found for brown capuchins' faces (one tailed paired t-test: $T = 2.27$, $df = 49$, $P = 0.01$) (Fig. 4), but not for faces of white-faced capuchins (one tailed paired $T = 0.24$, $df = 49$, $P = 0.4$), longtailed macaques (one tailed paired $T = 0.71$, $df = 49$, $P = 0.24$) or humans (one tailed paired $T = 0.091$, $df = 49$, $P = 0.46$). A species-specific limitation in the recognition of faces is also observed in brown capuchin monkeys.

## DISCUSSION

In both humans and non human primates tested in this study, we found the same limitation of recognition to own species faces. No evidence of recognition was found even for phylogenetically close species. These results support the hypothesis of a species-specific face recognition system in adults' primates. We observed that Tonkean macaques and brown capuchin monkeys looked longer at the pictures that presented their own species. This result is consistent with Fujita's study (1987) that showed that different species of macaques pressed a lever longer to see pictures of their own species compared with pictures of other macaque species.

Our results are however contradictory to other results from the literature that found that non human primates could process faces of other species. For example, Parr et al. (2000) showed that human-reared chimpanzees are able to discriminate between human faces as assessed in a forced choice task. The difference may reflect a methodology issue. The visual paired comparison task is a way to assess spontaneous recognition as it doesn't involve any training or reward. By contrast, the matching to sample paradigm requires extensive training and monkeys with poor recognition performance at the beginning could purposely used a special behavioral strategy to succeed, that is they may process the faces of other-species in a very different way than their own species faces (Ridley & Baker 1991; Pascalis & Bachevalier 1999). The same pattern of results is observed in humans who do not present evidence of discrimination of other species faces with the VPC but can do it in a forced choice task (Dufour et al. 2004). However when the stimulus' presentation time is short, humans succeed only at recognizing human faces in a forced choice task. Our results show that the face recognition system does not automatically process faces of another species.

There is a second explanation for the discrepancy observed between the results of this study and of Parr et al. (2000) study. In humans, early exposure influences face processing abilities (Pascalis et al. 2005); similar mechanisms could exist in non human primates. Our non human primate population was naturally reared by their mother in their own species social group and encountered only sometimes humans in their environment (keepers, primatologists). Such limited experience was insufficient to allow our non human primates to discriminate between human faces whereas extensive experience may have sculpted the chimpanzees face processing toward human faces. It would be consistent with Nelson's hypothesis (Nelson 2001) of an experience dependent face recognition system for non human primates.

The ability to discriminate at first sight between unknown individuals of its own species may represent a strong advantage when living in social groups, whereas categorizing individuals as not being from its own species is efficient enough. From an evolutionary point of view, such a system may be more economical than a system allowing discrimination between any faces of any primate species and could have been privileged by natural selection. Inability to spontaneously discriminate between individuals of another primate species could also prevent hybridization. The primates face recognition system demonstrates complex abilities and a high degree of specialization. It may have been an important tool toward the development of most primate societies that are based on strong social relationships (Thierry 1994).

The face discrimination limitation in primates may provide evidence that the common ancestor to actual living social primates needed such a specialized system to develop a social organization. In primates, the development of three-dimensional vision, with two eyes frontally oriented and the shortening of the "nose" (Simons 1963) occurred around 45 millions of years ago. In primate history, macaques separated 15 millions of years ago and capuchins lineage diverged between 45 and 28 millions of years ago (Kay et al. 2002). As we showed, this latter New World species has the same limitation in its face recognition system than other primates. Thus, we can assume that the selection of such a face recognition system will be anterior to the radiation of the capuchin ancestor from the primate order. The system would have kept comparable abilities throughout evolution. One way to confirm this hypothesis would be to study lemurs' face recognition ability. Lemurs separated earlier than capuchins inside the primate order, their communication relies more on olfactory than on visual cues (Gosset et al. 2003) and they do not show the "face like" pattern as simians do.

AKNOWLEDGMENTS

We thank Dr Bernard Thierry and Dr Jim Stone for useful comments on the manuscript. This work was supported by the Foundation Rotary International and the European Doctoral College of the Universities of Strasbourg, France.

REFERENCES

**Bahrick, H. P., Bahrick, P. O. & Wittlinger, R. P.** 1975. Fifty years of Memory for Names and Faces: a Cross-Sectional Approach. *Journal of experimental Psychology: General*, 104, 54-75.

**Bentin, S. Allison, T. Puce, A., Perez, E. & Mc Carthy, G.** 1996. Electrophysiological studies of faces Perception in Humans. *Journal of Cognitive Neuroscience*, 8, 551-565.

**Bruce, C.** 1982. Face recognition by monkeys: Absence of an inversion effect. *Neuropsychologia*, 20, 515-521.

**Campbell, R., Pascalis, O., Coleman, M., Wallace, S. B. & Benson, P. J.** 1997. Are faces of different species perceived categorically by human observers? *Proceedings of the Royal Society of London, B, Biological Science,* 264, 1429-1434.

**Campbell, R., Walker, J., Benson, P.J., Wallace, S., Coleman, M., Michelotti, J. & Baron-Cohen, S.** 1999. When does the inner-face advantage in familiar face recognition arise and why? *Visual Cognition,* 6, 197-216.

**Carey, S. & Diamond, R.** 1994. Are Faces Perceived as Configurations More by Adults than by Children ? *Visual Cognition,* 1, 253-274.

**de Haan, M., Pascalis, O. & Johnson, M. H.** 2002. Specialization of neural mecanisms underlying face recognition in human infants. *Journal of Cognitive Neuroscience*, 14, 199-209.

**Dasser, V.** 1987. Slides of Group Members as Representation of the Real Animals (*Macaca fascicularis*). *Ethology*, 76, 65-73.

**Diamond, R. & Carey, S.** 1986. Why faces are and are not special an effect of expertise. *Journal of Experimental Psychology, General,* 115, 107-117.

**Dufour, V. Coleman, M., Campbell, O., Petit, O & Pascalis O.** 2004. On the species-specificity of face recognition in human adults. *Current Psychology of Cognition*, 22, 315-333.

**Fujita, K.** 1987. Species recognition by five macaque monkeys. *Primates*, 28, 353-366.

364 **Fujita, K.** 1990. Species Preference By Infant Macaques with Controlled Social Experience.
365 *International Journal of Primatology,* 6, 553-573.

366 **Goren, C. C., Sarty, M. & Wu, P. Y. K.** 1975. Visual following and pattern discrimination of
367 face like stimuli by newborn infants. *Pediatrics*, 56, 544-549.

368 **Gosset, D., Fornasieri, I. & Roeder, J.** 2003. Acoustic structure and contexts of emission of
369 vocal signals by black lemurs. *Evolution of Communication*, 4, 225-251.

370 **Humphrey, N. K.** 1974. Species and Individuals in the Perceptual World of Monkeys.
371 *Perception,* 3, 105-114.

372 **Kay, R. F., Williams, B. A. & Anaya, F.** 2002. The Adaptation of Branisella Boliviana, the
373 Earliest South American Monkey. In *Reconstructing Behaviour in The Primate Fossil Record*
374 (Ed. by J. M. Plavcan, R. F. Kay, W. L. Jungers & C. P. van Schaik), pp. 339-370. New York:
375 Kluwer Academic/ Plenum.

376 **Kendrick, K. M., Atkins, L. L., Hinton, M. R.** 1998. Mothers determine sexual preferences.
377 *Nature*, 395, 229-230.

378 **Kendrick, K. M., Atkins, L. L., Hinton, M. R., Broad, K. D., Fabre-Nys, C. & Keverne, B.**
379 1995. Facial and Vocal Discrimination in Sheep. *Animal Behaviour*, 49, 1665-1676.

380 **Kendrick, K. M., Atkins, L. L., Hinton, M. R., Heavens, P. & Keverne, B.** 1996. Are faces
381 special for sheep ? Evidence of inversion effect and social familiarity. *Behavioural Processes,*
382 38, 19-35.

383 **Kendrick, K. N. da Costa, A. P., Leigh, A. E. Hinton, M. R. Pierce, J. W.** 2001. Sheep don't
384 forget a face. *Nature*. 441: 165–6.

385 **Kim, J. H., Gunderson, V. M. & Swartz, K. S.** 1999. Humans All Look Alike: Cross-species
386 Face Recognition in Infant Pigtailed Macaque Monkeys. *Poster for the Biennal Meeting of the*
387 *Society for Research in Child Development*, Albuquerque.

**Kuwahata, H., Adachi, I., Fujita, K., Tomonaga, M. & Matzuzawa, T.** 2004. Development of schematic face preference in macaque monkeys. *Behavioral Processes*, 66, 17-21. doi:10.1016/j.beproc.2003.11.002.

**Leder, H. & Bruce, V.** 2000. When Inverted Faces are Recognised: The role of configural information in face recognition. *The Quarterly Journal of Experimental Psychology,* 53, 113-536.

**Lutz, C. K., Lockard, J. S., Gunderson, V.M., Grant, K. S.** 1998. Infant Monkeys' Visual Responses to Drawings of Normal and Distorted Faces. *American Journal of primatology,* 44, 169-174.

**Maurer, D. Le Grand, R. & Mondloch, C. J.** 2002. The many faces of configural processing. *Trends in Cognitive Sciences*, 6, 255-260.

**Meissner, C. A. & Brigham, J. C.** 2001. Thirty years of investigating the own-race bias in memory for faces: A meta-analytic review. *Psychology, Public Policy and Law*, 7, 3-35.

**Myowa-Yamakoshi, M. & Tomonaga, M.** 2001. Development of face recognition in an infant gibbon (*Hylobates agilis*). *Infant Behavior and Development*, 24, 215-227. doi:10.1016/S0163-6383(01)00076-5.

**Nelson, C. A.** 2001. The development and neural bases of face recognition. *Infant and Child Development*, 10, 3-18. doi:10.1002/icd.239.

**Nelson, C. A.** 2003. The development of face recognition reflects an experience-expectant and activity-dependent process. In *The development of face processing in infancy and early childhood* ( Ed. by O. Pascalis & A. Slater)*,* pp. 77–98. New-York: Nova science publications.

**Nemanic, S., Alvarado, M. C., Bachevalier, J.** 2004. The hippocampal/parahippocampal regions and recognition memory: insights from visual paired comparison versus object-delayed nonmatching in monkeys. *Journal of Neuroscience*, 24, 2013-26. doi: 10.1523/jneurosci.3763-03.2004.

**Niedenthal, P.M., Halberstadt, J.B., Margolin, J. & Innes-Ker, A.H.** 2000. Emotional state and the detection of change in facial expression of emotion. *European Journal of Social Psychology*, 30, 211-222.

**Overman, W. H. & Doty, R. W.** 1982. Hemispheric specialization displayed by man but not by macaques for analysis of faces. *Neuropsychologia,* 20, 113-128.

**Parr, L.** 2003. The discrimination of faces and their emotional contents by chimpanzees (*Pan troglodytes*). *Annual New York Academy of Science*, 1000, 56-78. doi: 10.1196/annals.1280.005.

**Parr, L., Dove, T. & Hopkins, W. D.** 1998. Why faces may be special: Evidence of the inversion effect in Chimpanzees. *Journal of cognitive Neuroscience*, 10, 615-622.

**Parr, L., Winslow, J. T. & Hopkins, W. D.** 1999. Is the inversion effect in rhesus monkeys face-specific ? *Animal cognition,* 2, 123-129.

**Parr, L. A., Winslow, J. T., Hopkins, W. D. & de Waal, F. B. M.** 2000. Recognizing facial cues: Individual discrimination by chimpanzees (*Pan troglodytes*) and rhesus monkeys (*Macaca mulatta*). *Journal of Comparative Psychology*, 114, 47-60. doi: 10.1037//0735-7036.114.1.47

**Pascalis, O. & Bachevalier, J.** 1998. Face recognition in Primates: a cross-species study. *Behavioral Processes*, 43, 87-96.

**Pascalis, O. & de Haan, M.** 2003. Recognition Memory and Novelty Preference: What Model? In *Progress in Infancy Research* (Ed. by H. Hayne & J. W. Fagen), vol. 3, pp. 95-120. London, Mahwah: Lawrence Erlbaum Associates.

**Pascalis, O., de Haan, M. & Nelson, C. A.** 2002. Is Facing Processing Species Specific During the First Year of Life? *Science,* 296, 1321-1323. doi: 10.1126/science.1070223.

**Pascalis, O., Petit, O., kim, J. H. & Campbell, R.** 1999. Picture Perception in Primates: the case of Face Perception. *Current Psychology of Cognition,* 18, 889-921.

**Pascalis, O., Scott, L. S., Kelly, D. J., Shannon, R. W., Nicholson, E., Coleman, M. Nelson, C. A.** 2005. Plasticity of face processing in infancy. *Proceedings of the National Academy of Science*, 102, 5297-5300.

**Perrett, D. I., Hietanen, J. K., Oram, M. W. & Benson, P. J.** 1992. Organization and Function of Cells Responsive to Faces in the Temporal Cortex. *Philosophical Transaction of the Royal Society of London, B, Biological Sciences*, 335, 23-30.

**Phelps, M. T. & Roberts, W. A.** 1994. Memory for Pictures of Upright and Inverted Primates faces in Humans (*Homo sapiens*) Squirrel Monkeys (*Saimiri sciureus*) and Pigeons (*Colombia livia*). *Journal of Comparative Psychology*, 108, 114-125.

**Porter, R.H. & Bouissou, M.F.** 1999. Discriminative responsiveness by lambs to visual images of conspecifics. *Behavioural processes*, 48, 101-110.

**Ridley, R.M. and Baker, H.F.** 1991. A critical evaluation of primate models of amnesia and dementia. *Brain Research Reviews*, 16, 15-37.

**Simons, E. L.** 1963. A Critical Reappraisal of Tertiary Primates. In *Evolutionary and Genetic Biology of Primates* (Ed. by J. Buettner-Janusch), vol. 1, pp. 65-129. New York: Academic Press.

**Tanaka, M.** 2003. Visual preference by chimpanzees (Pan troglodytes) for photos of primates measured by a free choice-order task. Implication for influence of social experience. *Primates,* 44, 157-165. doi: 10.1007/s10329-002-0022-8.

**Thierry, B.** 1994. Emergence of Social Organizations in Non-Human Primates. *Revue Internationale de Systemique,* 8, 65-77.

**Tomonaga, M.** 1994. How laboratory-raised Japanese monkeys (*Macaca fuscata*) Perceice Rotated Photographs of Monkeys: Evidence for an Inversion Effect in Face Perception. *Primates*, 35, 155-165.

**Tomonaga, M., Itakura, S. & Matsuzawa, T.** (1993). Superiority of conspecific faces and reduced inversion effect in face perception by a chimpanzee. *Folia Primatologica,* 61, 110-114.

**Valentine, T.** 1991. A unified Account of the Effect of Distinctiveness, Inversion and Race in Face Recognition. *The Quaterly Journal of Experimental Psychology,* 43, 161-204.

**van Hoof, J. A. R. A. M.** 1967. The facial displays of the catarrhine monkeys and apes. In *Primate Ethology* (Ed. by D. Morris), pp. 7-68. London: Weindenfeld & Nicholson.

**Wright, A. A. & Roberts, W. A**. 1996. Monkey and Human Face Perception: Inversion Effects for Human Faces but not for Monkey Faces or Scenes. *Journal of Cognitive Neuroscience,* 8, 278-290.

**Yin, R. K.** 1969. Looking at upside-down faces. *Journal of Experimental psychology*, 81, 141-145.

**Zola, S. M., Squire, L. R., Teng, E., Stefanacci, L., Buffalo, E. A., Clark, R. E.** 2000. Impaired Reeognition Memory in Monkeys after Damage Limited to the hippocampal Region. *The Journal of Neuroscience*, 20, 451-463.

474  Figure 1: Main dates of radiation inside the primate order.

475  Figure 2: Illustration of the stimuli shown to participants. a: *Homo sapiens*, b: *Macaca*
476  *fascicularis*, c: *Macaca Tonkeana*, d: *Macaca mulatta*, e: *Macaca arctoides*, f: *Cebus apella*, g:
477  *Cebus capucinus*.

478  Figure 3: Illustration of gaze direction toward right (a) and left (b) stimuli in one capuchin. The
479  corneal reflection of the slide on the left or on the right part of the cornea indicates the direction
480  of looking.

481  Figure 4: Time spent looking toward the new and familiar stimulus for each species

482

482  1

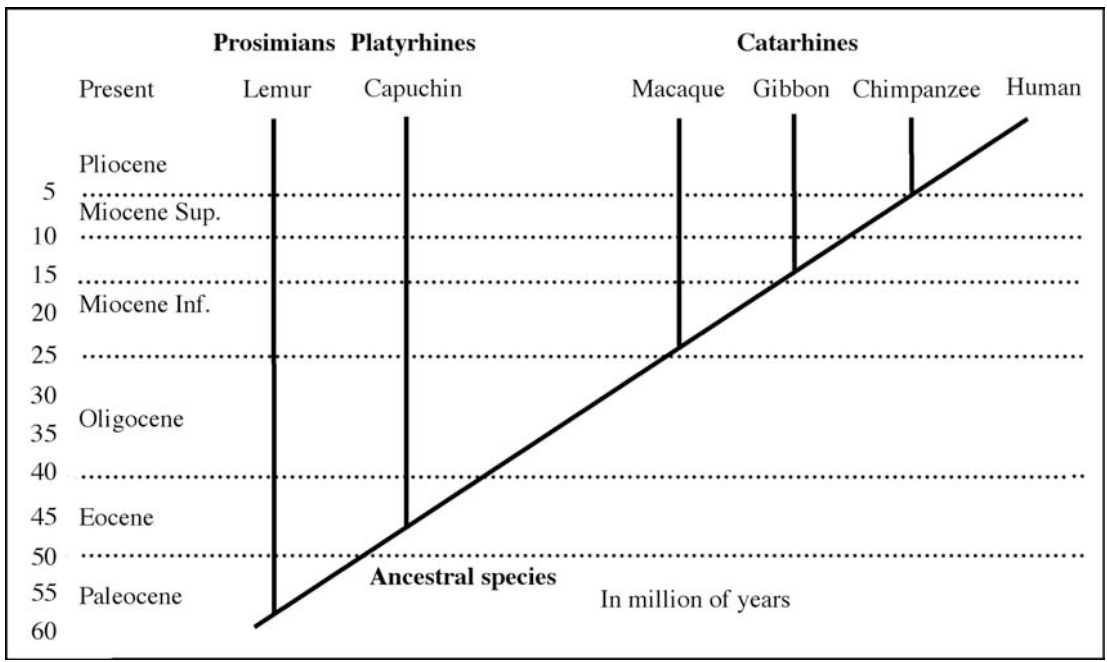

484    2

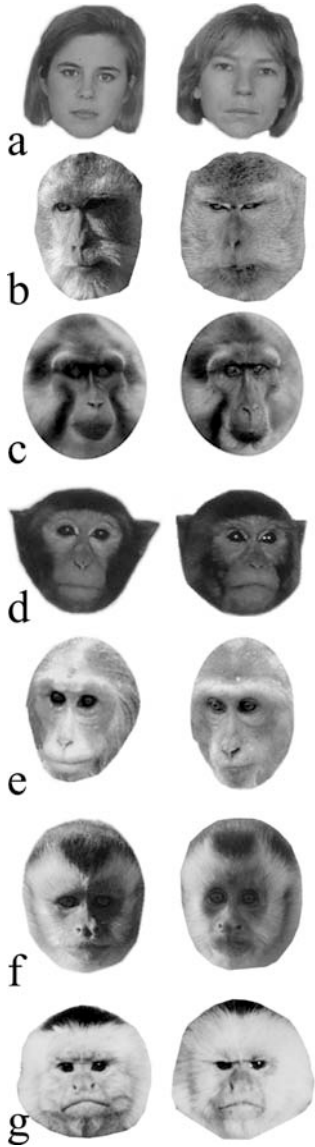

485    g

486

486  3

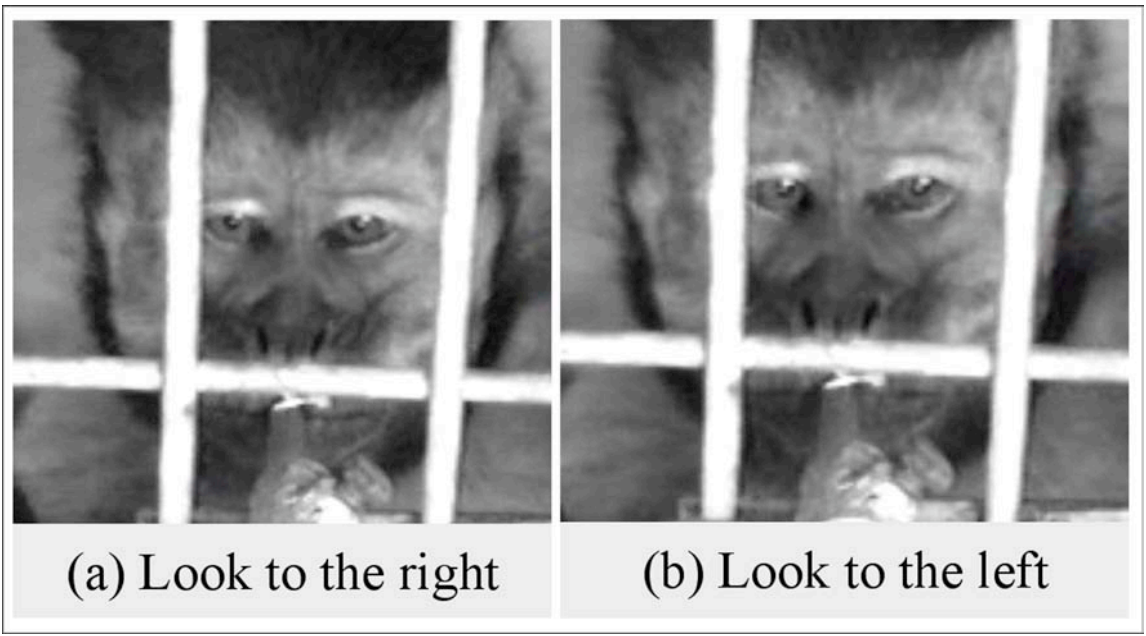

487

488  4

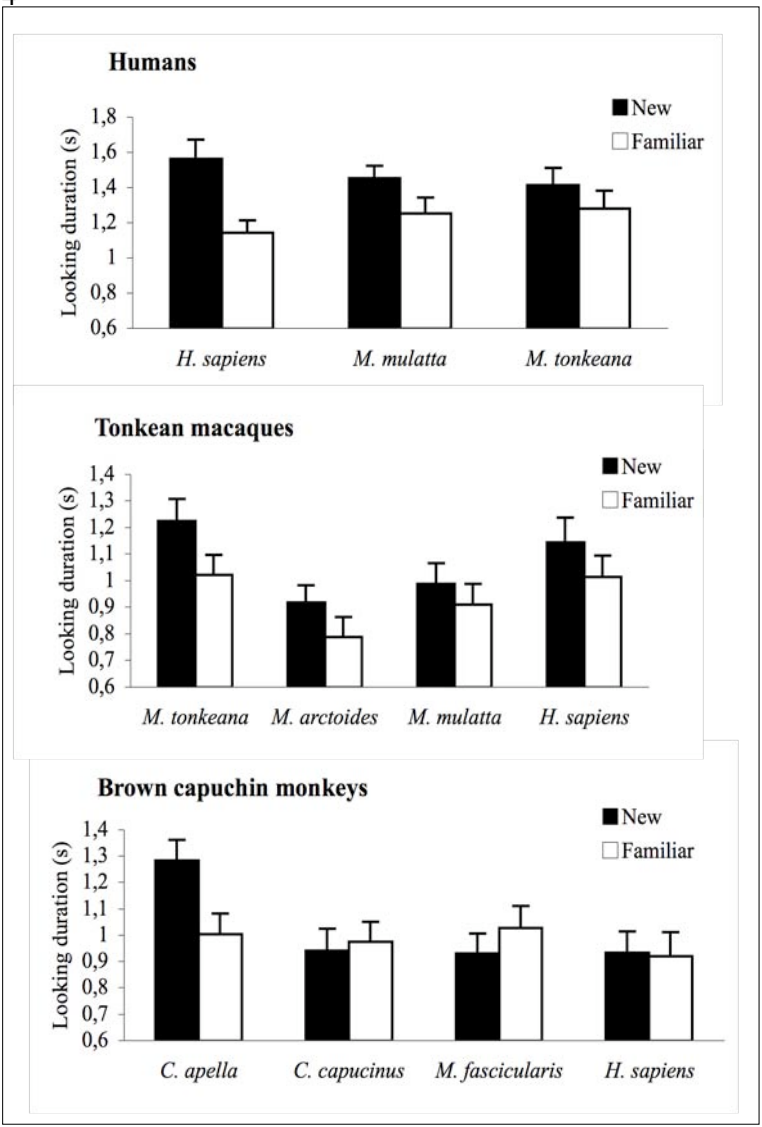